\documentclass{desyproc}

\usepackage{siunitx}
\usepackage{subfig}

\begin{document}
%------------------------------------
\title{An InGrid based Low Energy X-ray Detector}

%for single authors the superscripts are optional
\author{{\slshape Christoph~Krieger$^1$, Klaus~Desch$^1$, Jochen~Kaminski$^1$, Michael~Lupberger$^1$ and\\Theodoros~Vafeiadis$^{2}$}\\[1ex]
$^1$Physikalisches Institut, University of Bonn, Germany\\
$^2$CERN, Geneva, Switzerland}

% if the proceedings are available online (e.g. at Indico)
% please enter the contribution ID or file_name below for the DOI
%\contribID{32}
\contribID{krieger\_christoph}

% TO THE CONFERENCE EDITORS: 
% please update the following information      
% before sending the template to the authors
% \confID{800}  % if the conference is on Indico uncomment this line
\desyproc{DESY-PROC-2014-XX}
\acronym{Patras 2014} % if you want the Acronym in the page footer uncomment this line
\doi  % if there is an online version we will register DOIs

\maketitle

\begin{abstract}
An X-ray detector based on the combination of an integrated Micromegas stage with a pixel chip has been built in order to be installed at the \textbf{C}ERN \textbf{A}xion \textbf{S}olar \textbf{T}elescope. Due to its high granularity and spatial resolution this detector allows for a topological background suppression along with a detection threshold below $\SI{1}{\keV}$.

Tests at the CAST Detector Lab show the detector's ability to detect X-ray photons down to an energy as low as $\SI{277}{\eV}$. The first background data taken after the installation at the CAST experiment underline the detector's performance with an average background rate of $\SI[per-mode=repeated-symbol]{5e-5}{\per\keV\per\centi\metre\squared\per\second}$ between $\num{2}$ and $\SI{10}{\keV}$ when using a lead shielding.
\end{abstract}

\section{InGrid - An integrated Micromegas stage}

To enhance the performance of Micromegas detectors, it is necessary to match the granularity of the readout to the highly granular gas amplification stage. Taking into account the rising number of readout channels per area when following this approach, a pad based readout is impractical. This drawback can be bypassed by using integrated electronics in form of a pixel chip, e.g. the Timepix ASIC~\cite{llopart2007}. This pixel chip offers $\num{256}\times\num{256}$ pixels with a pixel pitch of $\SI{55}{\micro\metre}$ and thus an active area of $\SI{2}{\centi\metre\squared}$. Each of the pixels contains a charge sensitive amplifier and a discriminator plus complete counting logic needed for time or charge measurements.

\begin{figure}[tb]
\begin{center}
\begin{minipage}[b]{0.3\textwidth}
\subfloat[]{\label{subfig:ingrid}\includegraphics[trim = 0 45 0 0,clip=true,width=\textwidth]{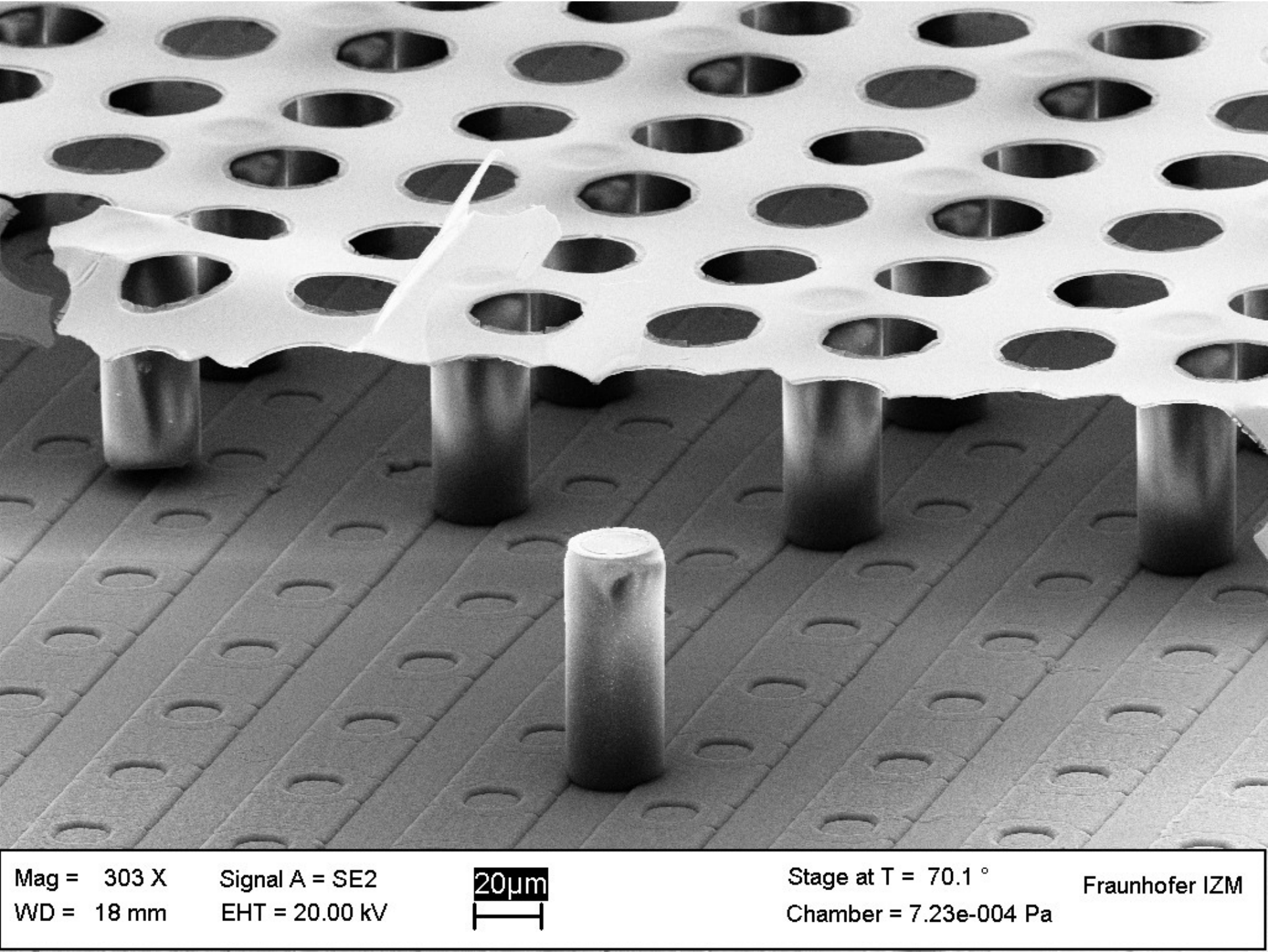}}\\
\subfloat[]{\label{subfig:sampleevent}\includegraphics[width=\textwidth]{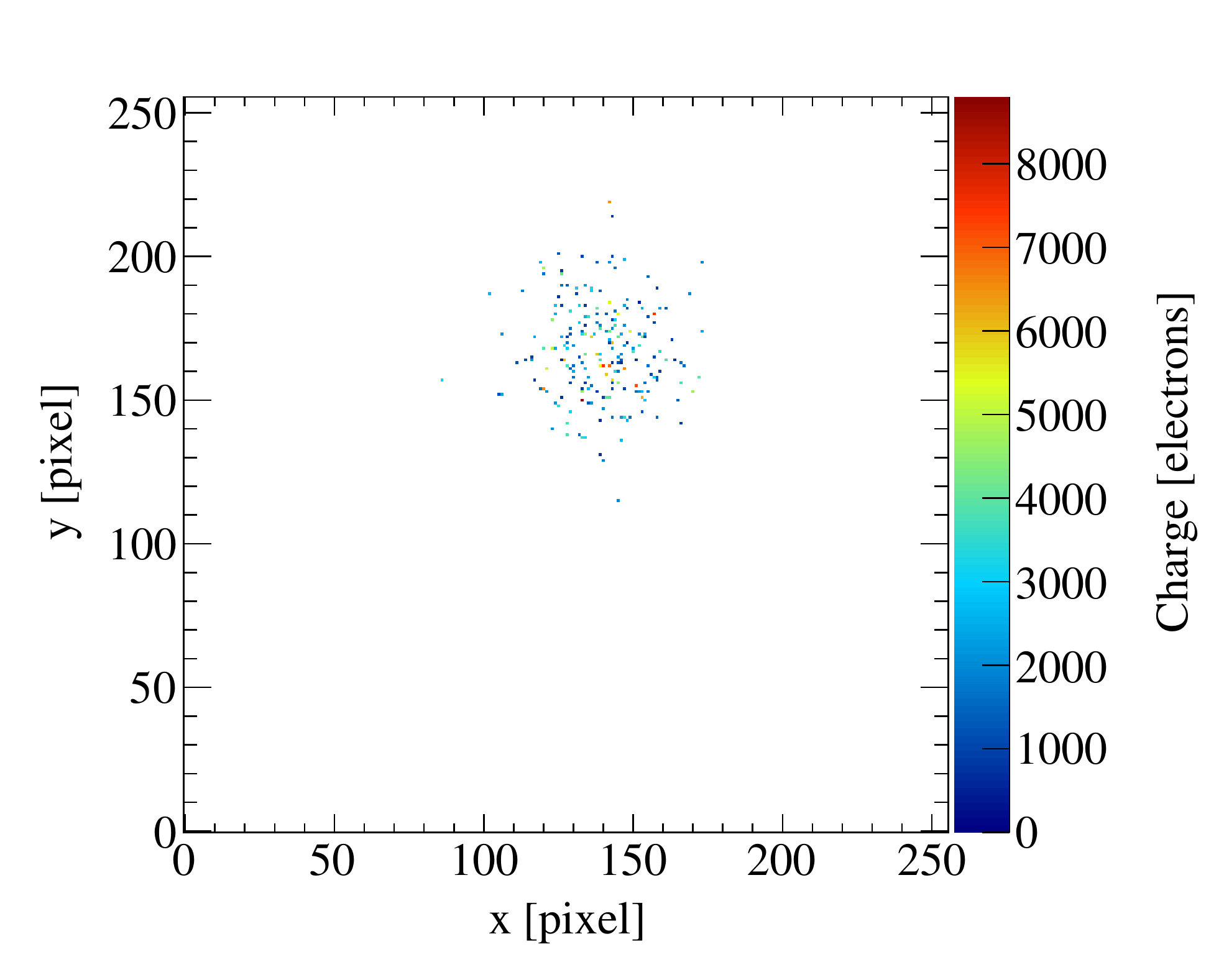}}
\end{minipage}
\hspace{.5cm}
\subfloat[]{\label{subfig:detector-ex}\includegraphics[trim = 200 460 160 120,clip=true,width=.6\textwidth]{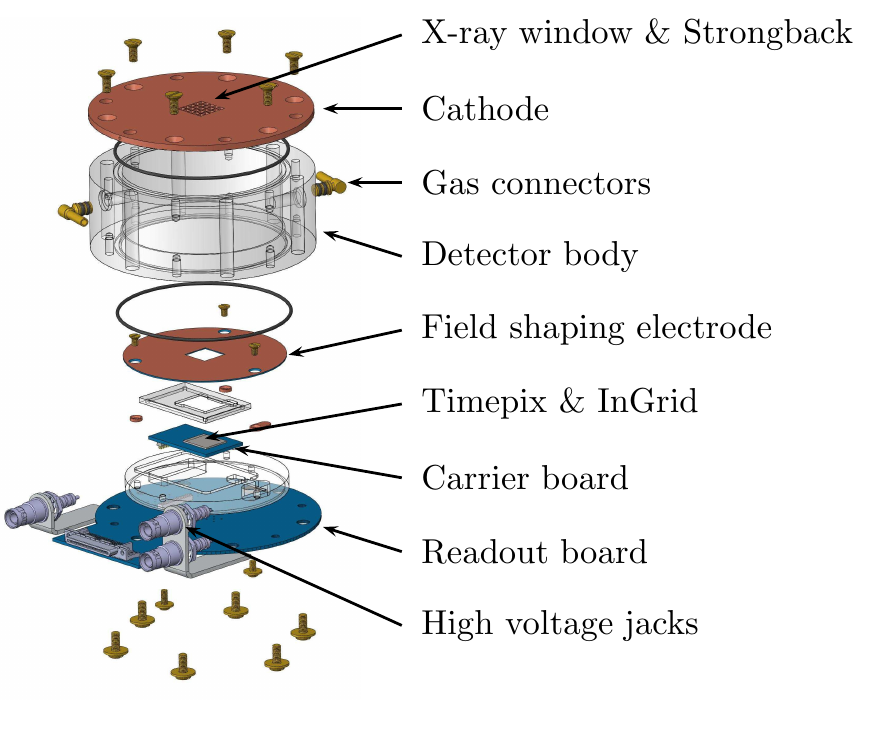}}
\end{center}
\caption{Scanning electron microscope picture of an InGrid structure \protect\subref{subfig:ingrid}, taken from~\cite{lupberger2014}. Event display showing an X-ray photon of $\SI{5.9}{\keV}$ from an $^{55}$Fe source as recorded with the InGrid based X-ray detector \protect\subref{subfig:sampleevent}, depicted area shows the complete active area of the Timepix ASIC. Z-axis shows the charge collected on each pixel. An exploded view of the detector, with main parts labeled, is shown in \protect\subref{subfig:detector-ex}. Drawing is taken from~\cite{krieger2014}.}
\label{fig:ingrid-detector}
\end{figure}

In order to achieve a precise alignment between the mesh holes and the pixels, to avoid the appearance of Moir\'e patterns, it is suitable to produce the Micromegas mesh directly on top of the pixel chip by means of photolithographic postproceesing technologies~\cite{chefdeville2006,vandergraaf2007}. A scanning electron microscope picture of the resulting structure can be seen in Figure~\ref{subfig:ingrid}. When building such an \textbf{In}tegrated \textbf{Grid} on top of a Timepix ASIC a resistive layer made of $\SI{4}{\micro\metre}$ silicon nitride is deposited between ASIC and InGrid to protect the chip and its electronics from discharges occurring during operation of the Micromegas stage~\cite{bilevych2011}.

\section{The InGrid based X-ray Detector}

The InGrid based X-ray detector constructed at Bonn uses a Timepix ASIC with an InGrid stage as central charge multiplication and readout device. The design of the detector (see Figure~\ref{subfig:detector-ex} for an exploded view) has been based on the recent Micromegas detectors~\cite{aune2009} used at the Sunset stations of the \textbf{C}ERN \textbf{A}xion \textbf{S}olar \textbf{T}elescope~\cite{zioutas1999}. The detector body is made of acrylic glass and contains the modular readout assembly which houses the Timepix ASIC, with the InGrid stage, mounted on a small carrier board. The chip is covered by a field shaping electrode, closing the readout assembly. The electrode features a cutout matching the size of the chip's active area and is leveled a bit above the mesh of the InGrid structure and set to the electric potential according to its position within the drift field. This helps to reduce electric field distortions arising at the chip's borders and the wire bonds connecting the chip's electronics to 
the 
underlying carrier board.

The drift volume featuring a drift distance of $\SI{3}{\centi\metre}$ at a drift field of $\SI[per-mode=symbol]{500}{\volt\per\centi\metre}$ is closed by a cathode plate made of copper. To allow especially low energy X-ray photons to enter the detector, a $\SI{2}{\micro\metre}$ thick aluminized Mylar film is used as entrance window. The material of the cathode plate above the instrumented area of the detector has therefore been removed except for a small grid like structure to support the thin window. The support for the window is necessary as it has to withstand a pressure difference of $\SI{1050}{\milli\bar}$. Additionally only a small leak rate is allowed so the detector can be operated connected to vacuum. Considering these requirements the choice of a $\SI{2}{\micro\metre}$ Mylar film represents a good compromise between robustness and transparency.

Readout of the Timepix ASIC is done with an FPGA based readout system developed at Bonn~\cite{lupberger2014} which allows for full access to firm- and software for customization. The detector is filled with a gas mixture composed of $\SI{97.7}{\percent}$ Argon and $\SI{2.3}{\percent}$ isobutane as quencher gas. 

The detection of X-rays with this detector is based on the fact that X-rays entering the detector will hit a gas atom and produce a bunch of primary electrons through ionization. These will drift towards the readout. The initial bunch of electrons is spread to a cloud of approximately circular shape due to diffusion. Then they get multiplied in the InGrid stage and are afterwards detected on the Timepix's pixels. This allows for a low X-ray energy threshold along with a topological background suppression by application of an event shape analysis. A typical X-ray event recorded with the detector is depicted in Figure~\ref{subfig:sampleevent}.

\section{Installation \& first background rates}

To replace the pnCCD detector behind the X-ray telescope~\cite{kuster2007} at one of the four detector stations of CAST, a vacuum system has been constructed and built which allows for differential pumping and provides the necessary safety interlocks. For the differential pumping a $\SI{0.9}{\micro\metre}$ thick Mylar window is used. In April and May of 2014 the vacuum system and the InGrid based detector were installed including a laser guided alignment with the X-ray telescope. A lead shielding designed and manufactured by our colleagues from the University of Zaragoza was added mid of May. In two months of operation no detector related problems occurred. 

\begin{figure}[tb]
\begin{center}
\subfloat[]{\label{subfig:c-spectrum}\includegraphics[width=.45\textwidth]{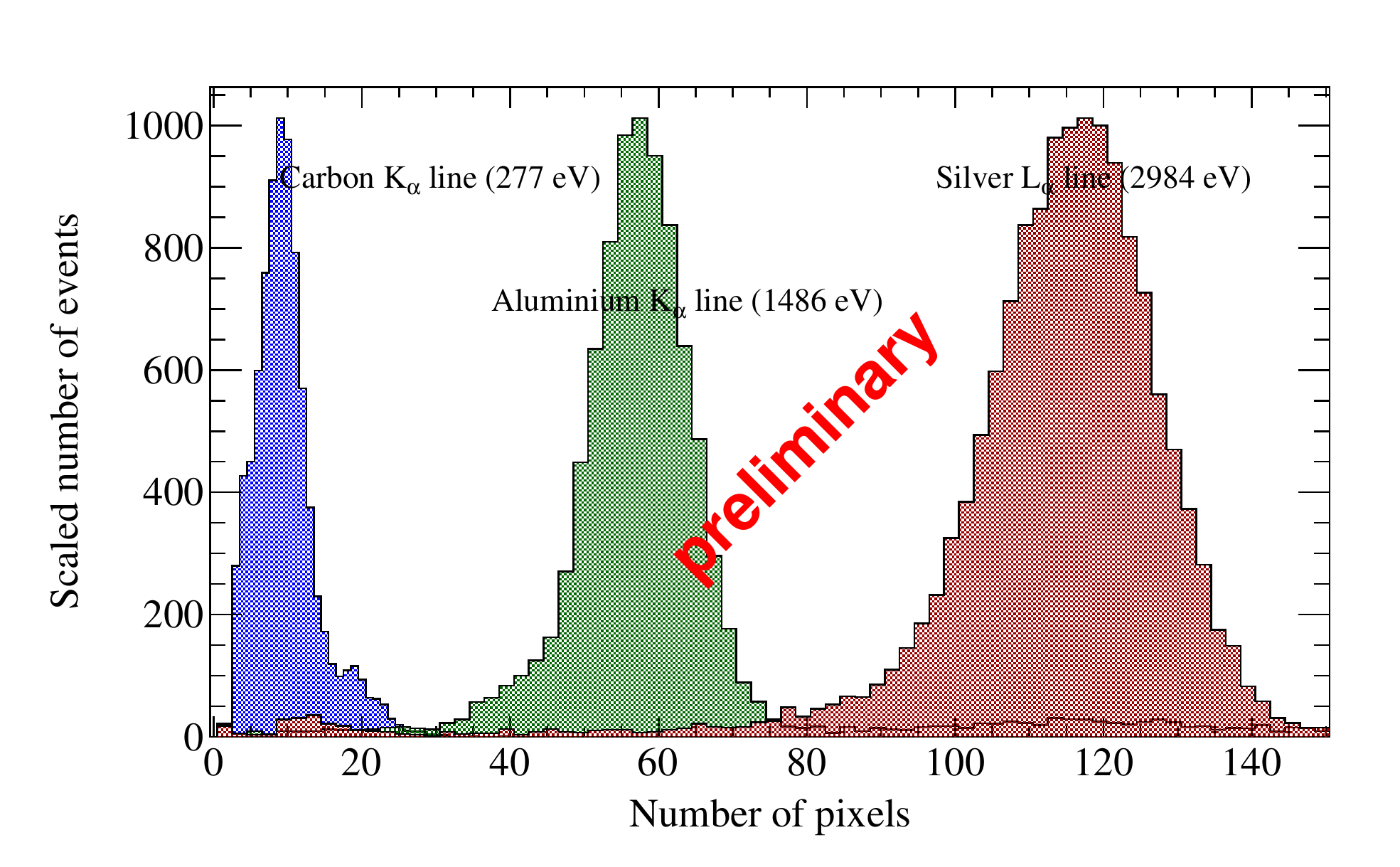}}
\hspace{.5cm}
\subfloat[]{\label{subfig:background}\includegraphics[width=.45\textwidth]{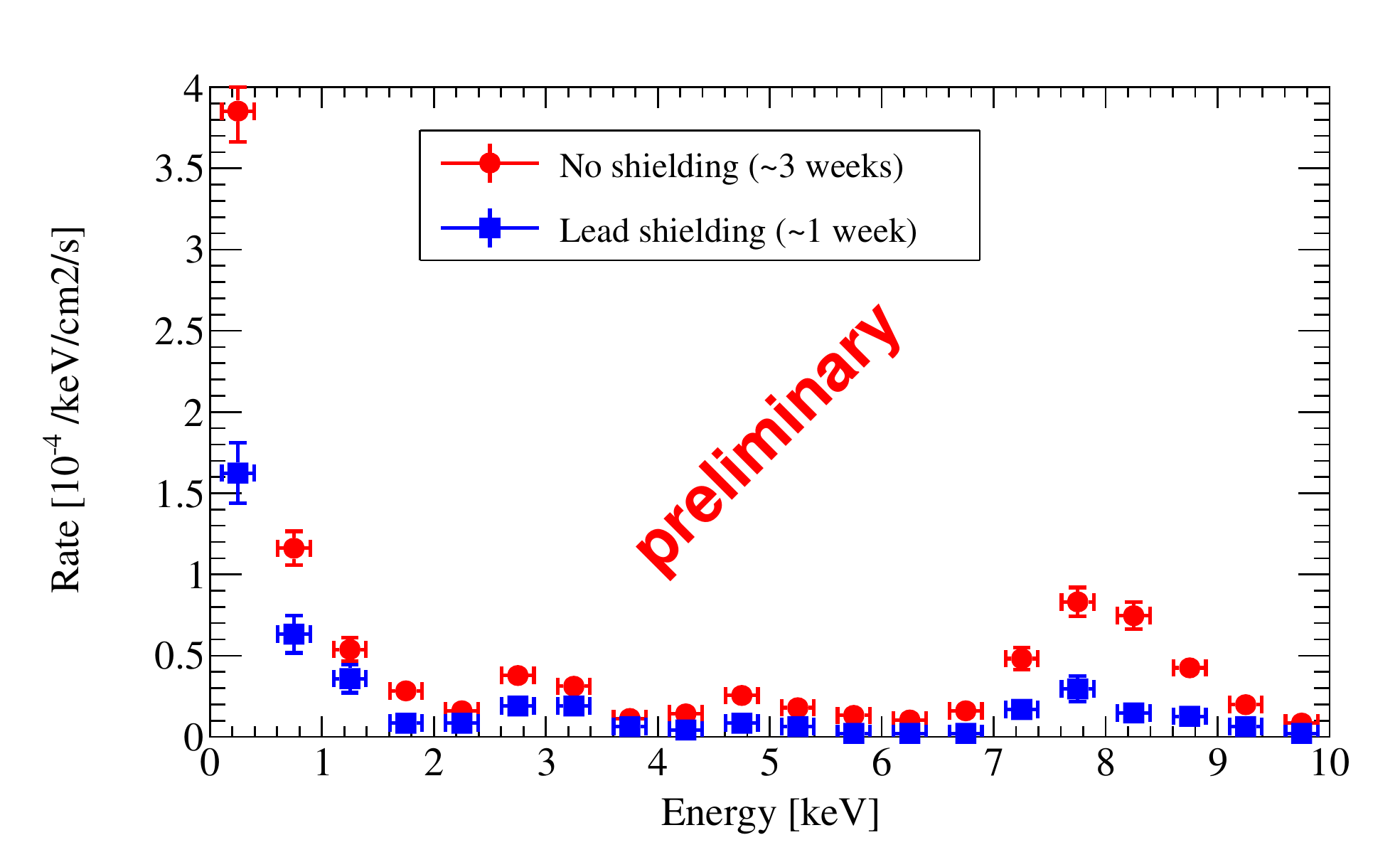}}
\end{center}
\caption{Spectrum showing the Carbon K$_\alpha$ line at $\SI{277}{\eV}$ together with the Aluminum K$_\alpha$ and the Silver L$_\alpha$ line at $\SI{1486}{\eV}$ and $\SI{2984}{\eV}$ respectively \protect\subref{subfig:c-spectrum}. Lines have been recorded separately. Number of pixels corresponds to the number of electrons in the charge cloud created by the initial X-ray photon which is proportional to the photon's energy. First background spectra obtained with and without lead shielding in the CAST environment \protect\subref{subfig:background}.}
\label{fig:c-spectrum-background}
\end{figure}

Prior to the installation, the detector was characterized at the X-ray generator of the CAST Detector Lab which provides a complete vacuum beamline connected to the X-ray generator~\cite{vafeiadis2012}. With the appropriate combination of targets of the X-ray tube and filters, characteristic X-ray lines can be produced in the energy range from a few hundred $\SI{}{\eV}$ up to $\SI{8}{\keV}$. Additionally the lab provides the necessary infrastructure for operating Micromegas detectors. The lab tests underlined the detector's low energy threshold through successful detection of photons down to the Carbon K$_\alpha$ line at $\SI{277}{\eV}$ (see Figure~\ref{subfig:c-spectrum} for the spectrum containing the Carbon K$_\alpha$ line).

To achieve a low background rate a background discrimination routine was created, based on a likelihood algorithm and reference data sets for different energy ranges recorded during the tests in the CAST Detector Lab. The applied likelihood method utilizes event shape properties making benefit of the Timepix ASIC's high spatial resolution resulting in a topological background suppression to identify real X-ray photons in the recorded data. The energy of a recorded event is calculated via a calibration curve from the total charge of an event. To built the likelihood, event shape properties of this event are being compared to the corresponding distributions for events of similar energy. These sample distributions are obtained from the lab tests' data. The likelihood cut value is adjusted for each energy range so that a signal efficiency of roughly $\SI{90}{\percent}$ is kept over all energies. The background rates achieved during first operation period were in the order of a few $\SI[per-mode=repeated-symbol]{e-5}{\per\keV\per\centi\metre\squared\per\second}$, occurring peaks could be explained by the characteristic fluorescence line of the copper cathode and misidentified cosmic rays traversing the detector perpendicular to the chip and therefore resulting in an X-ray like, circular, structure.

\section{Summary}

A gaseous detector based on the combination of a pixel chip and an integrated Micromegas stage has been successfully built and commissioned. The detector was installed at the CAST experiment along with its infrastructure and has up to now been successfully operated. First background rates achieved in the CAST environment look promising and are in agreement with the results obtained with a first prototype in the lab at Bonn~\cite{krieger2013}. Together with the measurements carried out at an X-ray generator facility which provided proof of the detector's low energy detection threshold of a few hundred $\SI{}{\eV}$, the detector's performance could be underlined and demonstrated.

Further improvement to the background discrimination algorithm will be done as well as future detector upgrades, which should include a decoupling and recording of the signal induced on the mesh, as it is done for the CAST Micromegas detectors. The latter one should give access to further event properties to be utilized for background suppression.
 
% ****************************************************************************
% BEGIN OF BIBLIOGRAPHY AREA
% ****************************************************************************

\begin{footnotesize}

\end{footnotesize}

% ****************************************************************************
% END OF BIBLIOGRAPHY AREA
% ****************************************************************************


\begin{thebibliography}{99}
\bibitem{llopart2007} X.~Llopart {\it et al.}, ``Timepix, a 65k programmable pixel readout chip for arrival time, energy and/or photon counting measurements'', Nucl.\ Instr.\ Meth.\ Phys.\ Res.\ A {\bf 581} (2007) pp 485-494.
\bibitem{chefdeville2006} M.~Chefdeville {\it et al.}, ``An electron-multiplying 'Micromegas' grid made in silicon wafer post-processing technology'', Nucl.\ Instr.\ Meth.\ Phys.\ Res.\ A {\bf 556} (2006) pp 490-494.
\bibitem{vandergraaf2007} H.~van~der~Graaf, ``GridPix: An integrated readout system for gaseous detectors with a pixel chip as anode'', Nucl.\ Instr.\ Meth.\ Phys.\ Res.\ A {\bf 580} (2007) pp 1023-1026.
\bibitem{bilevych2011} Y.~Bilevych {\it et al.}, ``Spark protection layers for CMOS pixel anode chips in MPGDs'', Nucl.\ Instr.\ Meth.\ Phys.\ Res.\ A {\bf 629} (2011) pp 66-73.
\bibitem{aune2009} S.~Aune {\it et al.}, ``New Micromegas detectors in the CAST experiment'', Nucl.\ Instr.\ Meth.\ Phys.\ Res.\ A {\bf 604} (2009) pp 15-19.
\bibitem{zioutas1999} K.~Zioutas {\it et al.}, ``A decommissioned LHC model magnet as an axion telescope'', Nucl.\ Instr.\ Meth.\ Phys.\ Res.\ A {\bf 425} (1999) pp 480-487.
\bibitem{lupberger2014} M.~Lupberger, ``The Pixel-TPC: first results from an 8-InGrid module'', J.\ Instrum. {\bf 9} (2014) C01033.
\bibitem{krieger2014} C.~Krieger {\it et al.}, ``An InGrid based Low Energy X-ray Detector for the CAST Experiment'', in proceedings of TIPP2014 conference, PoS(TIPP2014)060.
\bibitem{kuster2007} M.~Kuster {\it et al.}, ``The x-ray telescope of CAST'', New\ J.\ Phys. {\bf 9} (2007) 169 [arXiv: 0702188 [physics]].
\bibitem{vafeiadis2012} T.~Vafeiadis, ``Contribution to the search for solar axions in the CAST experiment'', CERN, Aristotle University, Thessaloniki 2012 [CERN-THESIS-2012-349].
\bibitem{krieger2013} C.~Krieger {\it et al.}, ``InGrid-based X-ray detector for low background searches'', Nucl.\ Instr.\ Meth.\ Phys.\ Res.\ A {\bf 729} (2013) pp 905-909.
\end{thebibliography}
\end{document}